DEC-TR-553

# Error Characteristics of FDDI

Raj Jain

Digital Equipment Corporation
550 King St. (LKG1-2/A19)
Littleton, MA 01460

Network Address: Jain@Erlang.DEC.Com

# Error Characteristics of Fiber Distributed Data Interface (FDDI)

Raj Jain
Digital Equipment Corporation
550 King St. (LKG1-2/A19)
Littleton, MA 01460

ABSTRACT

Fiber Distributed Data Interface (FDDI) is a 100 megabits per second fiber optic local area network (LAN) standard being developed by the American National Standard Institute (ANSI).

We analyze the impact of various design decisions on the error detection capability of the protocol. In particular, we quantify frame error rate, token loss rate, and undetected error rate. Several characteristics of the 32-bit frame check sequence (FCS) polynomial, which is also used in IEEE 802 LAN protocols, are discussed.

The standard uses a "non-return to zero invert on ones" (NRZI) signal encoding and a 4 bit-to-5 bit (4B/5B) symbol encoding in the physical layer. Due to the combination of NRZI and 4B/5B encoding, many noise events are detected by code (or symbol) violations. A large percentage of errors is also detected by framing violations. Some of the remaining errors are detected by FCS violations. The errors that escape these three violations remain undetected. The probability of undetected errors due to creation of false starting delimiters, false ending delimiters, or merging of two frames is analyzed.

It is shown that every noise event results in two code-bit errors, which in turn may result in up to four data-bit errors. The FCS can detect up to two noise events. Creation of a false starting delimiter or ending delimiter on a symbol boundary also requires two noise events. This assumes enhanced frame validity criteria. We justify the enhancements by quantifying their effect.

This analysis here is limited to noise events not resulting in a change of symbol boundaries. Extensions to the case of changed symbol boundaries is continuing and will be presented at a later time.

# 1  Introduction

The Fiber Distributed Data Interface (FDDI) is a 100 megabits per second ring network standard being developed by the American National Standard Institute (ANSI) [2,17]. The standard uses optical fibers as the transmission medium and allows rings with default maximum size of 1000 physical connections with a total fiber path length of up to 200 kilometers [1]. FDDI uses a timed token media access protocol proposed by Grow[4]. A number of papers have recently been published to analyze the performance and to prove certain operational characteristics of FDDI [1,9,18,20].

Optical fiber is known to have a lower bit error rate (BER) than the traditional copper wire. FDDI specifications require each fiber segment to have a bit error probability of less than 2.5E-10. The data encoding and frame formats have several reliability features that allow detection of errors and isolation of faults [11]. In particular, a "non-return to zero invert on ones" (NRZI) encoding is used to convert binary code-bits to optical pulses, five code-bits are combined to represent a symbol of four data-bits; and a frame check sequence (FCS) is used to check the integrity of the frame, which is delimited by a starting delimiter consisting of two control symbols and an ending delimiter consisting of one control symbol. This paper quantifies the combined impact of these design decisions on detected and undetected error rates.

We analyze the impact of noise in the optical signal on data-bits. It is shown that a single noise event may result in up to four data-bit errors. Several characteristics of the FCS polynomial are discussed. Undetected errors (UE) due to creation of a false starting delimiter, a false ending delimiter, or merging of two frames into one are analyzed.

The FDDI standards committee plans to enhance MAC specifications to improve the robustness of frame delineation in order to reduce the probability of undetected errors based on an earlier version of this analysis. A footnote describing these enhancement has already been added to MAC layer specifications (see p. 40 [2]). In this paper, we assume this enhanced version of MAC specifications. We also quantify the effect of these enhancements.

---
[1]These are *default maximum* values specified in the standard for the purpose of calculating default timer settings. Rings larger as well as smaller than the default size may be built. In the remainder of this paper, the term *maximum* when used with parameter values should be interpreted as the *default maximum* rather than as a limit.

# 2  FDDI Encodings

FDDI uses a serial baseband transmission system that combines the functions of data and clock transmission. Data recovery of this serial code-bit stream also provides recovery of synchronizing clock information.

The optical signals on the FDDI fibers use NRZI encoded pulses, where a polarity transition represents a logical "1" (one). The absence of a polarity transition denotes a logical "0" (zero). These logical ones and zeros are called *code-bits*. Five consecutive code-bits are grouped to form a *symbol*. Each symbol thus consists of five *code-bits*. The term *code cell* is used to denote the time interval of one code-bit. The receiving logic detects the changes in optical signal levels from one cell to the next.

The 5-bit symbols provide 32 possible bit combinations. As shown in Table 1, three of these symbols are reserved as line state symbols for use on the medium between frame transmissions; five symbols are used as control characters for frame delimiting and status indication; 16 symbols are used for data transmission within frame boundaries; and the remaining eight symbols are not used.

Detection of line state symbols (Quiet, Halt, and Idle) within a frame pre-empts and abnormally terminates any data transmission sequence in progress. Control symbols are named J, K, T, R, S. Each frame starts with a starting delimiter consisting of the two symbols JK and ends with an ending delimiter consisting of a T symbol. The frame also has a variable number of frame status indicators following the ending delimiter. Each of these status indicators can take only two values - set or reset. Symbols S and R are used to indicate set and reset, respectively.

A data symbol conveys one quartet (four data-bits) of arbitrary data within a frame. The elements of the 16 data symbols are denoted by the hexadecimal digits (0-F).

The code groups in 4B/5B encoding have been chosen so that during normal data transmission the DC component variation is less than $\pm 10\%$ from the nominal center [13]. There are at least two transitions per transmitted symbol and a transition in the optical signal occurs at least once every three cells, providing a cell-to-cell run length of three during frame transmission. Since edges (transitions) occur in the middle of a cell containing a one, the "edge-to-edge" run length is four. The bounded run length makes the signal self-clocking and simplifies clock recovery. Thus, symbols with three or more consecutive zero code-bits are not used as data symbols. The starting delimiter symbol pair JK has been chosen so that it will be recognized independent of the previously established symbol boundaries. In other words, code-bit sequence "1100010001" starts a new frame regardless of whether it occurs on a symbol boundary or not. The receiving logic of the physical layer

Table 1: 4B/5B Code

| Code Bits | Symbol | Assignment |
|---|---|---|
| Data Symbols: | | |
| 11110 | 0 | 0000 |
| 01001 | 1 | 0001 |
| 10100 | 2 | 0010 |
| 10101 | 3 | 0011 |
| 01010 | 4 | 0100 |
| 01011 | 5 | 0101 |
| 01110 | 6 | 0110 |
| 01111 | 7 | 0111 |
| 10010 | 8 | 1000 |
| 10011 | 9 | 1001 |
| 10110 | A | 1010 |
| 10111 | B | 1011 |
| 11010 | C | 1100 |
| 11011 | D | 1101 |
| 11100 | E | 1110 |
| 11101 | F | 1111 |
| Line State Symbols: | | |
| 00000 | Q | Quiet |
| 11111 | I | Idle |
| 00100 | H | Halt |
| Control Symbols: | | |
| 11000 | J | 1st of sequential SD pair |
| 10001 | K | 2nd of sequential SD pair |
| 01101 | T | Used to terminate the data stream |
| 00111 | R | Denoting logical zero (reset) |
| 11001 | S | Denoting logical one (set) |
| Invalid Code Assignments: | | |
| 00001 | VH | The code patterns marked V or VH |
| 00010 | VH | shall not be transmitted because |
| 00011 | V | they violate consecutive code-bit |
| 00101 | V | zeros or duty cycle requirements. |
| 00110 | V | Code marked VH shall however be |
| 01000 | VH | interpreted as Halt when |
| 01100 | V | received. |
| 10000 | VH | |

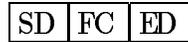

SD = Starting Delimiter (2 symbols)
FC = Frame Control (2 symbols)
ED = Ending Delimiter (2 symbols)

**Figure 1:** FDDI Token Format

(PHY) uses the incoming JK sequence to establish symbol boundaries.

The Halt symbol indicates a forced logical break in activity on the medium, while maintaining AC balance on the transmission medium. A continuous stream of Halt symbols is sent by a station to signal its presence on the outbound physical link to the neighboring station and to disable the associated physical connection without asserting control.

The Violation symbol V denotes a condition on the medium that does not conform to any other symbol in the symbol set. Violation symbols are not allowed to be transmitted onto the medium. The receipt of violation symbols may result from various error conditions or during ring clock synchronization sequences. The eight symbols listed as invalid code assignments are not allowed to be transmitted because they violate the run length or duty cycle requirements.

The IEEE 802.5 networks use a differential Manchester encoding scheme instead of the FDDI's 4B/5B with NRZI encoding. The differential Manchester encoding is rich in transitions, which simplifies the task of deriving the signal clock. However, it results in two pulses per data-bit and is, therefore, only 50% efficient. With Manchester encoding, the FDDI optical components and phase-locked loop would have to run at a signaling rate of 200 Mbaud. Instead, FDDI uses the 4B/5B encoding scheme, which is 80% efficient and requires only 125 Mbaud components [13].

# 3  FDDI Protocol Data Units

Two protocol data units (PDU) formats are used by FDDI MAC: tokens and frames. Each PDU is preceded by a preamble consisting of several Idle symbols. The size of the preamble varies as PDU travels around the ring and stations increase or reduce preamble to offset clock frequency differences from their upstream nodes. The remaining part of the token and frame formats are shown in Figures 1 and 2 respectively.

```
| SD | FC | DA | SA | INFO | FCS | ED | FS |
```

SD   = Starting Delimiter (2 symbols)
FC   = Frame Control (2 symbols)
DA   = Destination Address (4 or 12 symbols)
SA   = Source Address (4 or 12 symbols)
INFO = Information (0 or more symbol pairs)
FCS  = Frame Check Sequence (8 symbols)
ED   = Ending Delimiter (1 symbol)
FS   = Frame Status (3 or more symbols)

**Figure 2: FDDI Frame Format**

As shown in Figure 1, the token consists of a starting delimiter (SD), a frame control (FC) field, and an ending delimiter (ED). The starting delimiter is the symbol pair JK. The frame control field must be either 1000-0000 (non-restricted token) or 1100-0000 (restricted token). The non-restricted token is the normal token allowing asynchronous bandwidth to be time-sliced among all requesters. The restricted token allows all asynchronous bandwidth to be dedicated to a single extended dialog between specific requesters [2]. The ending delimiter for tokens consists of two T symbols.

The frame consists of a SD of two symbols JK, an FC of two symbols other than 1X00 0000, a destination address field of 4- or 12- symbols, a source address field of 4- or 12- symbols, INFO field of zero or more symbol pairs, a frame check sequence (FCS) of eight symbols, an ED of one T symbol, followed by three or more frame status (FS) indicator symbols. For details on interpretation of these fields, see FDDI MAC specifications [2].

The first three control indicators of the frame status field if present are used to indicate error detected (E), address recognized (A), and frame copied (C).

The E indicator is transmitted as R by the station that originates the frame. All stations on the ring inspect repeated frames for FCS errors. If an error is detected and the received E indicator was not Set, then an error is counted. The E indicator is set to S by a repeating station when an FCS error is detected in the frame.

The A indicator is transmitted as R by the station that originates the frame. If another station recognizes the destination address as its own individual or group address, it sets the A indicator to S; otherwise, a repeating station transmits this indicator as received.

The C indicator is transmitted as R by the station that originates the frame. If another station recognizes the destination address as its own and copies the frame into its receive buffer, it sets the C indicator to S; otherwise, a repeating station transmits this indicator as received.

## 4 Frame Validity Criteria

The analysis presented in this paper assumes the following enhanced frame validity criteria. A code-bit sequence is considered a valid frame if

1. It is a frame, i.e., it has a Starting Delimiter (JK), has an FC other than 1X00 0000, has zero or more additional data symbols, and has an ending delimiter (T). Here, X is either 0 or 1, r is reserved for standardization and should be set to zero.

2. It has a valid data length.

3. It has an FC=0X00 r000 or XX10 XXXX, or has correct FCS.

4. The ending delimiter (T) is followed by an E indicator with value R.

The first three criteria above are same as those stated in the standard ([2], p. 40). The fourth criteria in an enhancement which reduces the probability of a noise on one link validating a previously invalid frame.

Based on an earlier version of the analysis presented here, the standard committee has added a footnote to the standard ([2], p. 40) indicating its intent to enhance the frame validity criteria.

One implication of the above criteria is that each station on the ring should inspect the E indicator and handle it as follows:

1. If FC is neither 0X00 0000 nor XX10 XXXX and FCS is incorrect, set the E indicator.

2. Although stations on the ring can set the E indicator, they should *never* reset the indicator. This applies even if the FCS checks out OK.

3. If the E indicator is not R or S, it should be changed to S. This applies even if FCS is correct.

Later we will quantify the effect of these enhancements and show that the undetected error rates may not be acceptable without these.

# 5 Taxonomy, Notation, and Assumptions

In this section we define some of the terms used in the remainder of this paper.

We use the term *link* to denote all optical components from the transmit function of one PHY entity to the receive function of of the adjacent PHY entity. The link error rate includes errors in the fiber, connectors, optical receiver, and the optical transmitter.

As explained before, the FDDI uses a 4B/5B encoding to convert four data-bits to five code-bits. The code-bits are limited to the PHY layer. The media access control (MAC) layer deals only with symbols and data-bits. The term "bit" is used without a qualifier in this paper, if it is clear from the context whether it is data-bit or code-bit.

A *noise event* causes the receiver to misjudge the optical signal level, i.e., "on" (or high) may be interpreted as "off" (or low) and vice versa. We assume a non-bursty model for noise events, in that each event affects signal reception during only one code-cell duration. As we will see later, a single noise event results into two code-bit errors and one to four data-bit errors.

We use the following notation:

| | | |
|---|---|---|
| $L$ | = | Number of links in the ring |
| $l$ | = | Number of links between the source and destination of a frame |
| $p$ | = | Noise event probability per link (Link BER) |
| $F$ | = | Frame size in code-bits |
| $B$ | = | Link bandwidth in code-bits/second = 1.25E+8 for FDDI |
| $D$ | = | Ring latency |
| P(x) | = | Probability of event x |
| MT(x) | = | Mean time between events x |

FDDI standard specifies the following default maximum values of the ring parameters. The maximum number of links on the ring is 1000 ($L \leq 1000$). The maximum frame size is 9000 symbols ($F \leq 45000$ code-bits). The size includes four Idle symbols in the preamble and six control symbols in SD, ED, and FS fields ([2], section 4.3.5). The remaining symbols are data symbols. The maximum ring latency is 1.773 milliseconds. (The default maximum ring latency was changed from 1.617 milliseconds to 1.773 milliseconds in revision 15 of the PHY standard [3]). The maximum allowed *fiber link* bit error rate is 2.5E-10. This is the probability of noise events per link and should not be confused with code-bit or data-bit error probability which would be a multiple of this.

We make the following assumptions in the analysis presented here:

1. *Noise events are independent.* That is, occurrence of one noise event does not change the probability of occurrence of the next noise event. This simplifies the analysis considerably. This is a valid assumption if the noise is mostly due to thermal causes, which are independent in nature.

2. *Noise events are non-bursty.* That is, each event affects signal reception during only one code cell. As shown later, this results in bursty errors in the data-bits. Each noise event may result in a data error burst as long as four data-bits.

3. *The link can be modeled as a binary symmetric channel (BSC).* This means that the probability of a "high" level being interpreted as "low" on receipt is the same as that of a "low" signal being interpreted as "high".

4. *Noise events do not add or delete code-bits.* Only misinterpretation of signal levels are modeled. Addition or deletion of code-bits is left for future studies.

5. *The noise event probability $p$ is small.* Most expressions in this paper present only the lowest order term in $p$. Higher order terms make a negligible contribution if $p$ is small. This is not true if $p$ is close to 1. In general, we assume that $pLF \ll 1$, i.e., $p \ll \frac{1}{(45000)(1000)}$, or 1E-9.

6. *All data-bit patterns are equally likely.* In particular, this implies that all 16 data symbols (0-F) are equally likely in every data symbol position where data symbols are allowed.

7. *Data-bit errors in MAC layer electronic components are not modeled.* We consider only errors caused by misinterpretation of optical signal level. Electronic components, e.g., buses, memories, FCS logic, etc can cause errors in individual data-bits. Such errors are not modeled.

# 6 On Acceptable Error Rates

The maximum acceptable detected and undetected error rates vary not only among applications and environments but also with time. As the LAN technology is maturing, the minimum required reliability and data integrity is also increasing. Any specified numerical value of maximum acceptable error rates is bound to become outdated and even at the time of specifications it may not be applicable to some applications and environments. Nonetheless, it is important to set certain well specified goals to help select the design alternatives available at the time. This helps during the design phase in ruling out many alternatives that will not meet the goals. Also, it helps in setting configuration limits by ruling out the configurations that will not meet the requirements. For FDDI, this principle implies that the configuration limits (number of links per ring, length of the link, minimum
acceptable quality of links, etc.) and workload limits (frame length) should be chosen so
that the resulting performance, reliability, integrity, availability, and cost are acceptable.
In this paper, we are concerned solely with the error rates and want to ensure that the
error rates for any FDDI configuration and workload are reasonable.

Many transport protocols today are designed to allow a certain percentage of packet loss
due to congestion and errors. An end-to-end (over many hops) frame loss rate of 1%
is generally considered acceptable. A major part of this loss is allocated to congestion.
Thus, a fiber optic datalink with more than, say, 0.1% frame loss due to error alone may
be considered unacceptable. For unreliable media, such as radio links, one may either
allocate a larger share to error loss, or design higher level protocols to be able to sustain
a higher loss rate.

While the detected errors are harmful in that they require retransmissions resulting in
inefficient use of resources, undetected errors have no bounds on the damage that they
may cause. The damage caused by undetected errors in financial transactions or in defense
applications is unimaginable. One may, therefore, like to limit the number of undetected
errors per year to less than, say, 1/1000; that is, no more than one undetected error per
1000 years. For a manufacturer, this implies that if the manufacturer sells several thousand
FDDI networks, it will result in several undetected error cases per year, with each case
having a certain probability of resulting in a liability suit. For a user, such as a defense
installation, this implies that if the messages generally pass through, say, one hundred
LANs, the overall mean time between undetected errors will be about ten years.

The error analysis by nature tends to be pessimistic. This is because the designers want
to ensure an "upper bound" on errors. This is unlike traditional performance analysis
(throughput or delay analysis) in which "average" performance of an "average workload"
on an "average configuration" is more meaningful. For error analysis, one would like to
ensure that the error rates on all valid workloads (frame sizes and arrival rates) and on
all valid configurations (number of links, length of links, etc.) do not exceed a maximum
acceptable error rate. We, therefore, use the default maximum configurations (e.g., 1000
links, 4500 octet frames) as examples in this paper. Applications in which the resulting
error rates are unacceptable may further restrict allowable configurations or workloads.
We must point out though that the analysis presented here is not a 'worst case' analysis.
For example, we assume that all data symbols are equally likely. For a worst case scenario,
one could design frames consisting solely of symbols which are more likely to result in
undetected errors.

In the remainder of this paper, we use the term *large FDDI rings* to denote this default
maximum configuration with large size frames being continuously transmitted on the ring,
unless specified otherwise.

# 7  Effect of one noise event

Before we can compute the probabilities of detected and undetected errors in frames, we need to study the impact of a single noise event on a symbol in detail.

Consider the example of the symbol 0. It consists of four data-bits 0000 and using the 4B/5B coding, it is encoded into the five code-bits 11110, which in turn result in the transition sequence shown in Figure 3. A noise in the optical signal may cause the receiver to misjudge the signal level during the fourth code-cell, for instance, and so the received code-bit pattern is 11101, which is interpreted as symbol F, or data-bits 1111. This is an example of a single noise event resulting in four data-bit errors.

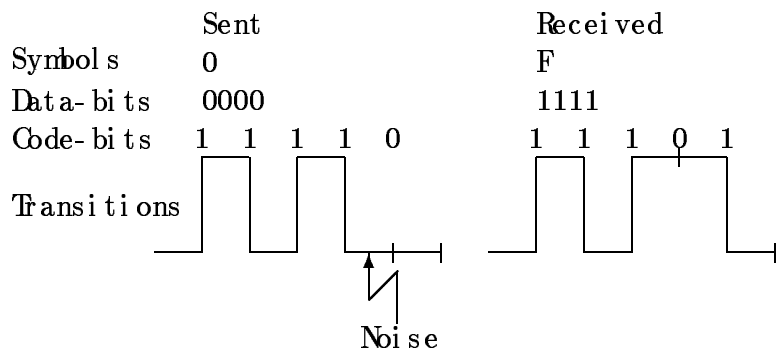

**Figure 3:** A single noise event can cause up to four data-bit errors.

The key observation from the above example is that *one noise event results in two code-bit errors*. This is true for all cases. If the noise affects the transition between two symbols, it affects the last (5th) code-bit of the first symbol, as well as the first code-bit of the second symbol.

Table 2 lists the effects of a noise on data symbols. Six possibilities are listed for each of the 16 data symbols. The first and the last column labeled code-bits 1 and 5 correspond to intersymbol errors, while the middle four columns are for intrasymbol errors. For example, the entry in the row labeled 3 and the column marked 4,5 is interpreted as follows. If the data symbol 3 (0011) is affected by noise so that its fourth and fifth code-bit positions are affected, the resulting symbol is A (1010).

From Table 2 we can compute the percentage of data symbol errors that result in other data symbols, control symbols, and violations. These percentages are listed in Table 3. The percentages for intrasymbol errors and intersymbol errors are given separately. The middle column labeled "count" in this table is simply the count of the resulting symbols

Table 2: Effect of Noise on a Data Symbol

| Original Symbol | Data-Bits | Code-Bits | Resulting Symbols Bit Positions Changed | | | | | |
|---|---|---|---|---|---|---|---|---|
| | | | 1 | 1,2 | 2,3 | 3,4 | 4,5 | 5 |
| 0 | 0000 | 11110 | 6 | V | 8 | J | F | I |
| 1 | 0001 | 01001 | S | K | V | 7 | 4 | VH |
| 2 | 0010 | 10100 | H | V | J | 8 | B | 3 |
| 3 | 0011 | 10101 | V | T | S | 9 | A | 2 |
| 4 | 0100 | 01010 | C | 8 | V | V | 1 | 5 |
| 5 | 0101 | 01011 | D | 9 | R | T | VH | 4 |
| 6 | 0110 | 01110 | 0 | A | VH | VH | T | 7 |
| 7 | 0111 | 01111 | I | B | V | 1 | V | 6 |
| 8 | 1000 | 10010 | VH | 4 | 0 | 2 | K | 9 |
| 9 | 1001 | 10011 | V | 5 | I | 3 | VH | 8 |
| A | 1010 | 10110 | V | 6 | C | VH | 3 | B |
| B | 1011 | 10111 | R | 7 | D | K | 2 | A |
| C | 1100 | 11010 | 4 | VH | A | E | S | D |
| D | 1101 | 11011 | 5 | V | B | F | J | C |
| E | 1110 | 11100 | V | H | VH | C | I | F |
| F | 1111 | 11101 | T | V | K | D | 0 | E |

in Table 2. For example, J occurs three times in the middle four columns (corresponding to the intrasymbol errors) of Table 2. Assuming each of the 16 data symbols is equally frequent, and that each of the five code cells is equally likely to be affected, this corresponds to $\frac{3}{(16)(5)}=3.75\%$

To study intersymbol errors, one needs to analyze all $(16)(16)=256$ data symbol pairs. The results of this analysis constitute the bottom half of Table 3.

In FDDI, many errors will be detected because the resulting code-bit pattern may translate to a *violation or invalid symbol*. The MAC layer keeps a count of format errors due to such symbol violations.

Some of the other errors will be detected if the resulting code-bit pattern translates to a *control symbol* which makes the frame an invalid frame, for example, a data frame ending with a symbol R rather than T. Such errors called *framing violations* are also counted by the MAC layer as format errors.

Table 3: Percentage of Data Symbol Errors

| Resulting Symbol | Count | Percent |
|---|---|---|
| Intrasymbol errors: | | |
| Data | 32 | 40.00% |
| J | 3 | 3.75% |
| K | 4 | 5.00% |
| R | 1 | 1.25% |
| S | 2 | 2.50% |
| T | 3 | 3.75% |
| H, I, V, VH | 19 | 23.75% |
| Subtotal | 64 | 80% |
| Intersymbol errors: | | |
| Data-data | 84 | 6.56% |
| Data-T | 14 | 1.09% |
| Data-R | 14 | 1.09% |
| Data-S | 14 | 1.09% |
| At least one H, I, V, VH | 130 | 10.16% |
| Subtotal | 256 | 20% |
| Total | | 100.00% |

Table 3 allows us to bound the probabilities of symbol violations and framing violations as follows:

1. 33.91% of the data errors result in I, V, or H symbols, which will cause the MAC layer to prematurely terminate the frame and replace the remaining part of the frame by Idle symbols. We call this *symbol violation*.

2. 46.56% of the data errors result in other data symbols and will not be detected by framing violations or symbol violations.

3. The remaining 19.53% of data errors result in control symbols which *may or may not* be detected by framing violations.

For those errors that result in new data symbols, it is interesting to analyze the data-bit error patterns. The results of this analysis are presented in Table 4. For each of the 16 data symbols, six possibilities are presented. A dash (-) is used to indicate the cases in which the resulting symbol is a nondata symbol. Notice that even though a single noise event can affect up to two symbols, it never affects more than four data-bits.

Notice from Table 4 that not all error patterns are equally likely. By counting the number of times an error pattern appears in this table we can compute the frequency of various error patterns. This is shown in Table 5. Again, intrasymbol and intersymbol errors have to be considered separately. For example, of the 256 possible data symbol pairs, 28 will result in a data-bit error pattern of 0001-0110, thereby, accounting for $\frac{28}{(256)(5)}=2.19\%$ of all data symbol errors. Notice that the sum of all data error pattern percentages is 46.56%, which is consistent with that in Table 3.

## 8 Frame Error Rate

A frame error results if the noise event affects any of the $F$ code cells in the frame. Also, a noise in the code cell immediately preceding the starting delimiter will affect the first code-bit of the frame. Given that each code-cell has a probability $p$ of being hit with noise, it is easy to compute the probability of no errors in any of the $F+1$ code-cells on any of the $L$ links.

$$\text{P(No error in } F+1 \text{ code-cells on any of the } L \text{ links)} = (1-p)^{L(F+1)}$$

$$\text{P(Frame error)} = 1 - (1-p)^{L(F+1)} \approx pLF \text{ for } pLF \ll 1$$

Table 4: Data Error Patterns

| Symbol | Error Pattern |||||| 
|---|---|---|---|---|---|---|
| | Bit Positions Changed |||||| 
| | 1 | 1,2 | 2,3 | 3,4 | 4,5 | 5 |
| 0 | 0110 | - | 1000 | - | 1111 | - |
| 1 | - | - | - | 0110 | 0101 | - |
| 2 | - | - | - | 1010 | 1001 | 0001 |
| 3 | - | - | - | 1010 | 1001 | 0001 |
| 4 | 1000 | 1100 | - | - | 0101 | 0001 |
| 5 | 1000 | 1100 | - | - | - | 0001 |
| 6 | 0110 | 1100 | - | - | - | 0001 |
| 7 | - | 1100 | - | 0110 | - | 0001 |
| 8 | - | 1100 | 1000 | 1010 | - | 0001 |
| 9 | - | 1100 | - | 1010 | - | 0001 |
| A | - | 1100 | 0110 | - | 1001 | 0001 |
| B | - | 1100 | 0110 | - | 1001 | 0001 |
| C | 1000 | - | 0110 | 0010 | - | 0001 |
| D | 1000 | - | 0110 | 0010 | - | 0001 |
| E | - | - | - | 0010 | - | 0001 |
| F | - | - | - | 0010 | 1111 | 0001 |

- $\Rightarrow$ The resulting symbol is a nondata symbol.

Table 5: Frequency of Data Error Patterns

| Error Pattern | Count | Percent |
|---|---|---|
| 0010 | 4 | 5.00% |
| 0101 | 2 | 2.50% |
| 0110 | 6 | 7.50% |
| 1000 | 2 | 2.50% |
| 1001 | 4 | 5.00% |
| 1010 | 4 | 5.00% |
| 1100 | 8 | 10.00% |
| 1111 | 2 | 2.50% |
| 0001-0110 | 28 | 2.19% |
| 0001-1000 | 56 | 4.38% |
| Total | | 46.56% |

The mean time between frame errors (sometimes referred to as error free seconds) can be computed if we know the mean time between frame arrivals. This time would be smallest on a fully utilized ring.

$$\text{Frames per second on a fully utilized link} \approx \frac{B}{F}$$

$$\text{Frames with error per second} = \frac{B}{F}\{1 - (1-p)^{L(F+1)}\}$$

$$\text{MT(Frame errors)} = \frac{1}{\frac{B}{F}\{1 - (1-p)^{L(F+1)}\}} \approx \frac{1}{BpL}$$

On large rings with large frames, the frame error probability comes out to 1.13% and the mean time between frame errors is 32 milliseconds. If this error probability is considered too high to be acceptable, the solution is to further restrict allowable values of $L$, $F$, or $p$. That is, decrease the number of links allowed on a ring, or decrease the maximum frame size allowed on the ring, or allow only higher quality components on the ring.

## 9 Token Loss Rate

As described earlier, the FDDI token consists of six symbols, i.e., 30 code cells. Error in any code cell or the code cell immediately preceding will cause the next station not to recognize the token resulting in a *token lost* event, which will eventually require the ring to be reinitialized with a new token. The probability of this event occurring during one pass around the ring can be computed in a manner similar to that for frame error rate with a frame size of $F = 30$ code-bits.

$$P(\text{Token loss per token rotation}) = 1 - (1-p)^{31L} \approx 31pL$$

On large rings the probability of token loss is 7.75E-6. On an idle ring, the token is continuously rotating around the ring. The mean time between token loss under such conditions can be computed as follows:

$$\text{MT(Token loss on an idle ring)} = \frac{\text{Ring latency}}{P(\text{Token loss per rotation})}$$
$$= \frac{D}{31pL}$$

For a large ring the ring latency is 1.773 milliseconds, which yields a mean time between token loss of 3.82 minutes. This is not the worst case time. For a given link BER, the time will be larger on busy rings and smaller on idle rings of smaller cable length. Since the ring latency is generally proportional to the number of links ($D \propto L$), the only way to increase this time (if unacceptable) is to allow only better quality links (with lower BER).

It should be pointed out that there are two types of tokens: restricted and non-restricted. These two types have been designed to differ from each other in only one *code-bit* position. Since a single noise event in the optical components always results in two code-bit errors, one event can not change a non-restricted token into a restricted token and vice versa.

# 10 FCS Polynomial

FDDI uses the following polynomial for the frame check sequence:

$$g(x) = x^{32} + x^{26} + x^{23} + x^{22} + x^{16} + x^{12} + x^{11} + x^{10} + x^{8} + x^{7} + x^{5} + x^{4} + x^{2} + x + 1$$

This polynomial is also used in IEEE 802 LAN standards [6,7,8] and in AUTODIN-II networks. For discussions related to errors in IEEE 802 protocols see references [12,15,19]. The polynomial was originally selected by Hammond et al [5] after comparing several 32 bit FCS polynomials listed in Peterson and Weldon's book [14]. It is listed there by its octal representation "40460216667".

One way to check if a frame has correct FCS would be as follows. Sequentially number the data-bits in the frame as $0, 1, 2, 3, \ldots$ starting with the data-bit before the ending delimiter and working backwards until the first data-bit after the starting delimiter. Let the $i^{th}$ data-bit be $b_i$, $b_i \in \{0,1\}$. The frame can then be represented by the polynomial:

$$f(x) = \sum_i b_i x^i$$

If the remainder $\text{Mod}(f(x), g(x))$ is zero, the frame is said to have the correct FCS[2]

This FCS polynomial has the following properties:

---
[2]This is a simplification. The FCS implementations as stated in the standards s condition:
$$\text{Mod}(x^n I(x) + x^{32}\{f(x) + I(x)\}, g(x)) = 0$$
Here, $n$ is the number of data-bits in the frame including FCS and $I(x) = \sum_{i=0}^{31} x^i$. The addition of I(x) in the above equation is equivalent to complementing the first 32 data-bits and the last 32 data-bits of the frame before the division operation.

1. It is a linear code. Linear codes have the important property that the "sum" of two code words is also a code word [14]. For FDDI and IEEE 802 protocols, this implies that if we take any two valid frames and do the following:

   (a) Right-align the frames,
   (b) Complement the first and the last 32 bits of each frame,
   (c) Take a bit-wise exclusive-or of their data-bits, and
   (d) Complement the first and the last 32 bits of the result

   The resulting data-bit sequence would form a frame with a valid FCS.

2. Adding a multiple of the divisor (FCS polynomial) to the dividend (frame polynomial) does not affect the remainder. The minimum degree polynomials, which are multiples of the FCS polynomial for various Hamming weights, are listed in Table 6 [16][3]. The Hamming weight of a polynomial is defined as the number of nonzero terms in the polynomial. For example, the $1 + x^{41678} + x^{91639}$ is a multiple of the FCS polynomial and has a Hamming weight of three. All other polynomials of lower degrees have higher weights. Such polynomials are important because if we add this polynomial to any frame, (this corresponds to complementing 0th, 41678th, and 91639th data-bits of the frame) the resulting FCS would still come out OK. Thus, for frames with lengths greater than or equal to 91640 data-bits (11455 octets), the minimum Hamming distance between two valid frames is three and the FCS can detect only two and one data-bit errors. Fortunately, this does not apply to FDDI or IEEE 802 since they do not allow such long frames.

3. For frames size between 3007 data-bits and 91639 data-bits, the minimum Hamming distance is four and the FCS detects all three, two, or one data-bit errors. This implies that for maximum size FDDI frames ($\approx$ 9000 symbols or 36000 data-bits), the FCS will not detect some four data-bit errors. Examples of four data-bit errors that will not be detected can be constructed by complementing the data-bits $i$, $i+2215$, $i+2866$, and $i+3006$ in any valid frame. This is true for all values of $i$.

   Similarly, statements can be made about other frame sizes by looking at the degree of polynomials in Table 6. The maximum frame size for various minimum Hamming distances are listed in Table 7 [16]. From this table we see that if the frame length is restricted to less than 375 octets, the minimum Hamming distance is five.

4. There are $2^d$ possible data-bit patterns which are $d$ data-bit long. Of these, only $2^{d-32}$ patterns have valid FCS. This is because given any data-bit pattern of $d-32$ we can compute its FCS and append it to make a valid $d$ data-bit pattern. Thus, the probability of any randomly constructed $d$ data-bit pattern to have a valid FCS is $\frac{2^{d-32}}{2^d}$ or $2^{-32}$ or 2.33E-10.

---

[3]The polynomials corresponding to Hamming weight of 8 and 11 were incorrect in [16]. The polynomials given in Table 6 are correct.

Table 6: Multiples of FCS Polynomial

| Hamming Weight | Minimum Degree Polynomial |
|---:|---|
| 3 | $1+x^{41678}+x^{91639}$ |
| 4 | $1+x^{2215}+x^{2866}+x^{3006}$ |
| 5 | $1+x^{89}+x^{117}+x^{155}+x^{300}$ |
| 6 | $1+x^{79}+x^{85}+x^{123}+x^{186}+x^{203}$ |
| 7 | $1+x^{45}+x^{53}+x^{74}+x^{80}+x^{120}+x^{123}$ |
| 8 | $1+x^{5}+x^{13}+x^{16}+x^{36}+x^{41}+x^{88}+x^{89}$ |
| 9 | $1+x^{2}+x^{3}+x^{18}+x^{19}+x^{32}+x^{37}+x^{57}+x^{66}$ |
| 10 | $1+x^{3}+x^{7}+x^{25}+x^{27}+x^{30}+x^{33}+x^{36}+x^{38}+x^{53}$ |
| 11 | $1+x^{5}+x^{7}+x^{16}+x^{31}+x^{32}+x^{35}+x^{37}+x^{41}+x^{43}+x^{44}$ |
| 12 | $1+x^{3}+x^{5}+x^{7}+x^{8}+x^{13}+x^{18}+x^{21}+x^{24}+x^{26}+x^{30}+x^{42}$ |
| 13 | $1+x+x^{6}+x^{15}+x^{18}+x^{20}+x^{23}+x^{29}+x^{33}+x^{35}+x^{37}+x^{40}+x^{42}$ |

Table 7: Hamming Distance of FCS Polynomial

| Hamming Weight | Max Frame Size Data-Bits | Octets |
|---:|---:|---:|
| 3 | 91639 | 11454 |
| 4 | 3006 | 375 |
| 5 | 300 | 37 |
| 6 | 203 | 25 |
| 7 | 123 | 15 |
| 8 | 89 | 11 |
| 9 | 66 | 8 |
| 10 | 53 | 6 |
| 11 | 44 | 5 |
| 12 | 42 | 5 |
| 13 | 42 | 5 |

5. If there are several data-bits in error in a frame, the group of data-bits beginning from the first data-bit in error up to the last data-bit in error is called an *error burst*. The burst size $b$ includes the first and the last data-bits (which are in error) and all intermediate data-bits (which may or may not be in error). The FCS polynomial detects *all* error bursts of size 32 or less. Thus, if several noise events affect a frame such that the resulting error burst is less than 32 data-bits, the FCS will detect it. The fraction of error bursts larger than 33 data-bits that are not detected is $2^{-32}$. For bursts of size exactly 33 data-bits, this fraction is $2^{-31}$ [14].

This property implies that all single noise events will be detected by the FCS since the event would produces a burst of at most four data bits.

The above statements do not say anything about two noise events that affect symbols far apart. One may suspect that some two noise events will not be detected by the FCS. Fortunately, this is not so. We know from the previous section, that there are only ten possible error patterns. An exhaustive search using a computer program showed that the FCS polynomial detects all possible two noise events. Some combinations of three noise events are not detected. For example, if we sequentially number the symbol positions of an FDDI frame as 0, 1, 2, ... starting from the last symbol position of the FCS field and proceeding backwards toward the FC field, and we introduce error patterns 1010, 1111, and 0010 in positions $i$, $i+625$, $i+3605$, respectively, the resulting frame will still have a valid FCS for all values of $i$. A complete list of other possible three noise events that will not be detected is shown in Table 8. The search included the possibility that a symbol may be affected by more than one noise events.

Also listed in the table are the corresponding probabilities. For example, to compute the probability corresponding to the first line of the table, we observe that only 5% of the data errors result in error pattern 1010, 2.5% of data errors result in the error pattern 1111, and 5% of the data errors result in the pattern 0010. A frame has $(F-50)/5$ data symbols, therefore, $0 \leq i \leq (\frac{F-50}{5} - 3605)$. The symbol error probability is $5p$. Assuming that there are $L/2$ links on an average between the source and destination, the required probability is:

P(Positions $i$, $i+625$, $i+3605$ are affected by error

$$\text{patterns 1010, 1111, and 0010 respectively}) = \sum_{\forall i}(0.05 \times 5p)(0.025 \times 5p)(0.05 \times 5p)(0.5L)$$

$$= \sum_{\forall i}(7.8125\text{E-}3)p^3(0.5L)$$

$$= (\frac{F-50}{5} - 3605)(7.8125\text{E-}3)p^3(0.5L)$$

Table 1: Complete List of Three Noise Events not Detected by the FDDI FCS

| Noise 1 | | Noise 2 | | Noise 3 | | Probability |
|---|---|---|---|---|---|---|
| Symbol Position | Error Pattern | Symbol Position | Error Pattern | Symbol Position | Error Pattern | For Large Rings |
| $i$ | 1010 | $i+625$ | 1111 | $i+3605$ | 0010 | 3.29E-25 |
| $i$ | 1000 | $i+1366$ | 1001 | $i+6398$ | 0010 | 1.58E-25 |
| $i$ | 1001 | $i+1630$ | 1001 | $i+5509$ | 1000 | 2.12E-25 |
| $i$ | 1111 | $i+1835$ | 1001 | $i+8404$ | 0101 | 1.79E-26 |
| $i$ | 0010 | $i+1947$ | 1111 | $i+3096$ | 1000 | 1.80E-25 |
| $i$ | 1100 | $i+2239$ | 00010110 | $i+3289$ | 0110 | 9.14E-25 |
| $i$ | 0101 | $i+3881$ | 00011000 | $i+5609$ | 0110 | 2.71E-25 |
| $i$ | 1100 | $i+3882$ | 0010 | $i+5609$ | 1000 | 4.13E-25 |
| $i$ | 00011000 | $i+4209$ | 1111 | $i+8972$ | 00010110 | 3.98E-28 |
| $i$ | 1001 | $i+6092$ | 0110 | $i+6340$ | 0101 | 2.43E-25 |
| | | | | | Total | 2.74E-24 |

$$= (\frac{F-50}{5} - 3605)(3.91\text{E-}3)\overset{3}{p}L$$

The total probability of undetected errors is obtained by summing it for all possible patterns listed in the table. For the largest size frames this probability is 2.74E-24. For other frame sizes the probability is approximately ( 3.89E$^3$$\overset{3}{p}$).p

Using the computer program, we also tried to prepare a table of four noise events that will not be detected. The table became too large much before reaching completion. The incomplete part did verify the theoretical argument that the fraction of undetected four noise events is $\overset{-}{2}^2$. This being so, the probability of undetected errors due to four noise events can be computed as follows:

P(Four noise events not causing

$$\text{symbol or FCS violations}) = \binom{F-50}{4}(0.4656p)^4(1-p)^{(F-50-4)}(L/2)(2^{-32})$$

$$\approx \frac{\{0.4656p(F-50)^4\}(L/2)(2^{-32})}{24}$$

$$\approx (2.28\text{E-}13)\overset{4}{p}F^4L$$

$$\approx (2.28\text{E-}13)\overset{4}{p}F^4L$$

On large rings with large frames, the probability of undetected errors due to four noise



Table 9: Maximum Frame Size vs Detected Noise Events on FDDI

| # of Noise Events | Maximum Frame Size | | | |
|---|---|---|---|---|
| | Symbols | | | Octets |
| | Data | Non-data | Total | |
| 3 | 3096 | 10 | 3106 | 1553 |
| 4 | 434 | 10 | 444 | 222 |
| 5 | 30 | 10 | 40 | 20 |

events is 3.64E-30. Probabilities for larger number of noise events can be calculated similarly.

The relationship between maximum frame size and the maximum number of noise events per frame allowed on FDDI is shown in Table 9. From this table we see that if the frame size is limited to 3106 symbols (3096 data symbols, four Idle symbols in the preamble, and six control symbols for the delimiters and status indicators), the FCS will detect all three noise events. For frames shorter than 444 symbols, the FCS will detect all four noise events. The corresponding number for five noise events is 40 symbols.

## 11 Merging Frames

On a dual (counter-rotating) ring, dual attachment stations connect to both (primary and secondary) rings. Some of these dual stations, called concentrators, may offer additional attachment points for other stations. The dual stations and concentrators can internally reconfigure their data paths to allow stations to be added to the ring or to be removed from the ring. If a station is allowed to go on/off the ring improperly, frames or parts of frames on the fiber connecting the station to/from the concentrator may be lost. It is possible to lose parts of two frames such that the resulting data-bit pattern is a valid frame as shown in Figure 4. Since the FCS is 32 data-bits long, the probability that any data-bit pattern has a valid FCS is $2^{-32}$ or 2.33E-10 or one in 4.34E+9. In other words, one in every 4.34 billion merged frames will have a correct FCS. This may or may not be acceptable depending upon the frequency of stations going on/off the ring and the number of stations. To avoid frame merging, it is recommended that the switching be done only during idle line states or that a format error be forced on incomplete frames every time a station goes on/off the ring.



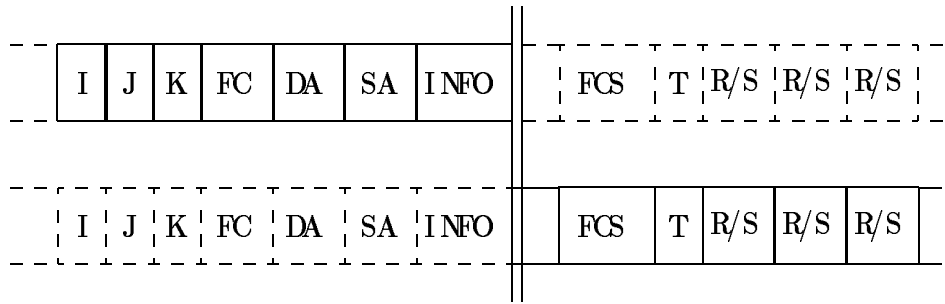

Figure 4: Two frames may merge to produce one valid frame.

## 12  False Ending Delimiter

FDDI uses a frame-ending delimiter of a single symbol T. However, with enhanced frame validity criteria, the T symbol must be followed by an E indicator with value R. Thus, we need at least two noise events changing two data symbols to a TR pair and create a false ending delimiter. If we examine this in more detail we find that data symbols changing to TR result in three possible scenarios:

1. The T appears in FC, DA, or SA fields. This is counted as a framing violation. The fraction of such *false T* is $\frac{130}{(F-50)}$, where 130 code-bits (13 octets assuming 12 symbol addresses) of the total $F - 50$ code-bits constitute these fields. The remaining 50 code-bits are used by the preamble, SD, ED, and FS fields.

2. The T appears in the second symbol of an octet in the INFO field. This results in an odd number of data symbols between SD and ED. This is also counted as frame violation. This fraction is $\frac{(F-180)}{(2F)}$. This approximates to about 50%

3. The T appears in the first symbol of an octet in the INFO field. This will result in a premature termination of the frame. Again, this fraction is $\frac{(F-180)}{2F}$. In other words, about half of the errors converting a data symbol to T will not be detected by framing violations.

It is also possible that for some of these frames with false ED, the FCS checks out OK. The probability of this is a product of the probability of the following events:

1. A noise event affects a data symbol.

2. The data symbol is the first symbol of an octet.



3. the data symbol becomes a T.

4. Another noise affects the next data symbol.

5. The second data symbol becomes an R.

6. FCS is correct.

The probability of the second event is 0.5. That of the sixth event is $2^{-32}$. The probability of the third event is 4.84% (sum of 3.75% and 1.09% in Table 3), and that of the fifth event is 1.25% (note that intersymbol errors result in R only if the previous symbol becomes a data symbol, hence they are not added in this probability). Thus,

$$
\begin{aligned}
P(\text{UE due to false ED}) &= (P(\text{a data symbol in odd position becoming T})) \\
&\quad (P(\text{FCS OK})) \\
&\quad (P(\text{the next data symbol becoming an R})) \\
&= (0.0484 \times 5p)\left(\frac{F-180}{5}\right)\left(\frac{1}{2}\right)\left(\frac{L}{2}\right)(2^{-32})(0.0125 \times 5p) \\
&\approx (1.76\text{E-}13)\, p^2 LF
\end{aligned}
$$

and

$$
M(\text{UE due to false ED}) = \frac{1}{\frac{B}{F}(1.76\text{E-}13)\, p^2 LF} = \frac{1}{(1.76\text{E-}13)\, Bp^2 L}
$$

For large rings and large frames, the probability of undetected errors is 4.93E-25 and the mean time between undetected errors is 2.31E+13 years. This is acceptable for most applications.

## 13  False Starting Delimiter

In FDDI, each frame starts with a JK symbol pair. It is possible to have two or more noise events so that we get a valid starting delimiter. Using the percentages specified in Table 3 and following a methodology similar to that for the false ED, we can compute the probability of undetected errors due to false SD as follows:

$$
P(\text{UE due to false SD}) = (0.0375 \times 5p)(0.05 \times 5p)\left(\frac{L}{2}\right)\left(\frac{F-180}{5}\right)\left(\frac{1}{2}\right)(2^{-32}) \approx (5.46\text{E-}13)\, LFp^2
$$

and

$$
M(\text{UE due to false SD}) = \frac{1}{(5.46\text{E-}13)\, BLp^2}
$$



For large rings and large frames, this probability is 1.53E-24 and the mean time between
undetected errors is 7.47E+12 years. This may be considered acceptable for most appli-
cations. Further, the starting delimiter is actually stronger than this since some of the
frames considered valid in the above analysis will have nonexistent destination addresses
and invalid frame control fields.

The analysis presented above assumes that JK code-bit pattern '1100010001' appears on
a symbol boundary. It does not account for cases in which the pattern may appear at
non-boundary positions. FDDI PHY layer will recognize such non-boundary JK's and
establish a new symbol boundary for the remaining stream. Analysis of such cases is
currently underway and will be reported elsewhere [10]. It should be pointed out though
that such non-boundary cases can be caused by a single noise event and are much more
likely than boundary cases analyzed here.

# 14 Need for Enhanced Validity Criteria

The analysis presented so far assumed enhanced frame validity criteria and frame-status
indicator handling rules. In this section we quantify the effect of these enhancements and
justify their need.

In general, the enhancements guarantee that all noise events required to create an unde-
tected errors must all appear on the same link. This is because if the noise events happen
on two different links, the errors will be detected by the station at the end of the first link
and the frame will be marked invalid with E indicator set. It is not possible for a single
noise event to change S to an R.

If E indicator is not mandatory, the ending delimiter would consist of a single symbol T.
A single noise event can change a data symbol to T and potentially cause the frame to end
prematurely.

$$P(\text{UE due to false ED w/o enhancements}) = (0.0484 \times 5p) \left(\frac{L}{2}\right) \left(\frac{F-180}{5}\right) \left(\frac{1}{2}\right) (2^{-32})$$
$$\approx (2.82\text{E-}12) pLF$$

and
$$M(\text{UE due to false ED w/o enhancements}) = \frac{1}{(2.82\text{E-}12) pLB}$$

For large rings and large frames, the probability is 3.16E-14. This may be considered
unacceptable for some applications.



Without the enhancements, the formula for undetected errors due to FCS would also be different. Without the enhancements, E indicator is not mandatory. It is possible for a frame without the E indicator to be affected by noise events on three different links such that after the third event the frame has a correct FCS and thus results in an undetected error. Assuming that there are $l$ links between the source and destination, the probability of a single error is $pl$ and that of three errors is $p^3l^3$. Assuming all values of $l$ between 1 and $L-1$ are equally likely, the average probability of three noise events would be $p^3\frac{L^3}{4}$ (since average of $1^3, 2^3, 3^3, \ldots, (L-1)^3$ is approximately $\frac{1}{4}L^3$). The approximate expression for probability of undetected errors due to three noise events is $(1.95E-13)pF^3L^3$. Thus, the enhancements improve this by a factor of $0.5L$.

Similarly, the average of $1^4, 2^4, 3^4, \ldots, (L-1)^4$ is approximately $\frac{1}{5}L^4$. The expression for probability of undetected errors due to four noise events without enhancements would be $(9.12E-14)pF^4L^4$. The enhancements improve this by a factor of $0.4L$.

# 15 Other (optional) Enhancements

The principal reason for originally making all status indicators optional was that some implementations of FDDI MACs may save costs by not checking the frame status indicators. However, if the E indicator becomes mandatory, the implementations may check the next two frame status indicators A and C as well. The incremental complexity to do this is small. Let us first analyze the impact of making the A indicator mandatory.

## 15.1 Option 1: A Indicator Must Be R or S

This option would require that frame sending and receiving stations will treat a frame as invalid whose A indicator is not R or S. In other words, if A indicator is *not available* it will be treated the same way as if the E indicator was set. The A indicator is not reset or set by any station. The receiving station sets the A indicator if and only if it is an R.

A indicator checking is not an alternative to E indicator checking. We assume that this option would be considered only if the E indicator checking has already been implemented. Implementing this option further reduces the probability of false ending delimiters. At least three noise events are required to create a valid ending delimiter. From Table 3, we find that the probability of getting an R/S from data symbols is $1.25+2.5=3.75\%$. Note that intersymbol errors can result in R/S only if the previous symbol becomes a data symbol,



hence they are not added in the above probability.

$$P(\text{UE due to false ED with option 1}) = (0.0484 \times 5p)\left(\frac{F-180}{5}\right)\binom{F}{2}\binom{L}{2}(2^{-32})$$
$$(0.0125 \times 5p)(0.0375 \times 5p)$$
$$\approx (3.30\text{E-}14)\hat{p}^3 LF$$

and

$$M(\text{UE due to false ED with option 1}) = \frac{1}{\frac{B}{F}(3.30\text{E-}14)\hat{p}^3 LF}$$
$$= \frac{1}{(3.30\text{E-}14)\hat{p}^3 BL}$$

For large rings, the probability is 2.32E-35 and the mean time between undetected errors is 4.92E+23 years. This rule reduces the undetected error rate by a factor of 4.70E-11.

It must be pointed out that this option makes the ending delimiter stronger than the FCS and the starting delimiter. Thus, even though, the probability of undetected errors due to false ending delimiter decreases considerably, the net undetected error rate remains close to that due to false FCS or due to a false starting delimiter and does not change. A indicator checking at the destination should, therefore, be optional rather than a requirement.

### 15.2 Option 2: C Indicator Must Be R or S

This rule would further strengthen the ending delimiter by requiring that if C indicator is not R or S, the frame be treated as invalid. The effect of this is similar to that of the previous option, i.e., the net gain of this rule is 4.69E-11. Again, this reduces undetected errors due to false ending delimiter but does nothing to the total undetected error rate as that is now governed by the FCS and the false starting delimiter and therefore, this rule should also be optional.

## 16 Summary

We have quantified the impact of various encoding and frame format decisions for FDDI. In particular, the impact of NRZI encoding, with 4B/5B coding, FCS polynomial, starting delimiter JK, ending delimiter T, and optional frame status indicators on the undetected error rates was analyzed in detail. The numerical results for 4500 octet frames on large
FDDI rings with 1000 links each with a noise event probability (BER) of 2.5E-10 are
summarized in Table 10. By changing each of the three key parameters, namely, noise
event probability, number of links, and frame size, by a factor of 10 and recomputing
the results as shown in table 10, we can get a sense of sensitivity of the results to these
parameters.

The results of this analysis are as follows:

1. A single noise event that results in misjudging the optical signal level during one
   code cell always results in two code-bit errors. This may result in one or two symbol
   errors and up to four data-bit errors.

2. For large rings, the frame loss rate or token loss rate may be too high for some
   applications and therefore it may be preferable to use higher quality links, a smaller
   number of stations, or shorter frames.

3. Several characteristics of the FCS polynomial were investigated and it was deter-
   mined that it detects all one or two noise events and that some three noise events
   may not be detected by the polynomial. For frames of 1553 octets or shorter it can
   protect against all three noise events.

4. A false starting delimiter of JK can be generated (on a symbol boundary) by two
   noise events.

5. A false ending delimiter of TR can be generated by two noise events.

6. If E indicator is not mandatory and if stations are allowed to reset the E indicator,
   the undetected error rates due to false ending delimiter may be unacceptable for
   some applications.

7. The A- and C- indicators may also be optionally checked. However, it does not
   decrease the total probability of undetected errors.

## 17   Acknowledgments

We would like to thank several people who helped during this analysis: Bob Grow, Bill
Hawe, Jerry Hutchison, Charlie Kaufman, Berry Spinney, Don Knudson, Paul Koning,
Tony Lauck, Dave Neuman, Bruce Thompson, Lih Wng, and Henry Yang. We are also
grateful to Mr. James Hamstra and another referee for several useful comments on an
earlier draft of this paper.

Table 2: Summary of Error Rates For FDDI Rings *

| Quantity | Unit | Large Rings | BER= 2.5E-11 | 100 Links | 450 Octets Frame |
|---|---|---|---|---|---|
| P(Frame error) | | 1.13E-02 | 1.13E-03 | 1.13E-03 | 1.13E-03 |
| MT(Frame error) | ms | 32. | 320. | 320. | 32. |
| P(Token loss per token rotation) | | 7.75E-06 | 7.75E-07 | 7.75E-07 | 7.75E-06 |
| MT(Token loss on an idle ring) | sec | 229. | 2288. | 229. | 229. |
| P(FCS not detecting 3 noise events) | | 2.74E-24 | 2.74E-27 | 2.74E-25 | 0[†] |
| MT(FCS not detecting 3 noise events) | year | 4.17E+12 | 4.17E+15 | 4.17E+13 | ∞ |
| P(FCS not detecting 4 noise events) | | 3.64E-30 | 3.64E-34 | 3.64E-31 | 3.49E-34 |
| MT(FCS not detecting 4 noise events) | year | 3.14E+18 | 3.14E+22 | 3.14E+19 | 3.27E+21 |
| P(UE due to false ED) | | 4.93E-25 | 4.93E-27 | 4.93E-26 | 4.75E-26 |
| MT(UE due to false ED) | year | 2.31E+13 | 2.31E+15 | 2.31E+14 | 2.40E+13 |
| P(UE due to false SD) | | 1.53E-24 | 1.53E-26 | 1.53E-25 | 1.47E-25 |
| MT(UE due to false SD) | year | 7.47E+12 | 7.47E+14 | 7.47E+13 | 7.75E+12 |
| Without enhanced E indicator handling rules: | | | | | |
| P(UE due to false ED) | | 3.16E-14 | 3.16E-15 | 3.16E-15 | 3.04E-15 |
| MT(UE due to false ED) | year | 362. | 3616. | 3616. | 375. |
| P(FCS not detecting 3 noise events) | | 1.37E-18 | 1.37E-21 | 1.37E-21 | 0[†] |
| MT(FCS not detecting 3 noise events) | year | 8.34E+06 | 8.34E+09 | 8.34E+09 | ∞ |
| P(FCS not detecting 4 noise events) | | 1.45E-21 | 1.45E-25 | 1.45E-25 | 1.40E-25 |
| MT(FCS not detecting 4 noise events) | year | 7.85E+09 | 7.85E+13 | 7.85E+13 | 8.17E+12 |
| With optional A indicator handling rules: | | | | | |
| P(UE due to false ED) | | 2.32E-35 | 2.32E-38 | 2.32E-36 | 2.30E-36 |
| MT(UE due to false ED) | year | 4.92E+23 | 4.92E+26 | 4.92E+24 | 4.97E+23 |

*Parameters if unspecified are: 1000 Links, BER=2.5E-10, 4500 octets frames.
[†] FCS detects all 3 noise events for frames shorter than 1553 octets.
Notation: P(.)=Probability of, MT(.)=Mean time between, UE=Undetected Error